\documentclass{ws-procs975x65}

\def\gtrsim{>}

\def\beq{\begin{equation}}
\def\eeq{\end{equation}}

\def\beqa{\begin{eqnarray}}
\def\eeqa{\end{eqnarray}}

\begin{document}

\title{COSMIC DAWN: STUDIES OF THE EARLIEST GALAXIES AND THEIR ROLE IN COSMIC REIONIZATION}

\author{R. S. ELLIS$^*$}

\address{Department of Astronomy, California Institute of Technology,\\
Pasadena, CA 91125, USA\\
$^*$E-mail: rse@astro.caltech.edu\\
www.astro.caltech.edu/$\sim$rse/}

\begin{abstract}
I review recent progress and challenges in studies of the earliest galaxies, seen when the Universe was less than 1 billion
years old. Can they be used as reliable tracers of the physics of cosmic reionization thereby complementing other, more
direct, probes of the evolving neutrality of the intergalactic medium? Were star-forming galaxies the primary agent in
the reionization process and what are the future prospects for identifying the earliest systems devoid of chemical
enrichment? Ambitious future facilities are under construction for exploring galaxies and the intergalactic medium
in the redshift range 6 to 20, corresponding to what we now consider the heart of the reionization era. I review 
what we can infer about this period from current observations and in the near-future with existing facilities, and 
conclude with a list of key issues where future work is required.
\end{abstract}


\bodymatter

\section{Introduction}

Most would agree that the final frontier in piecing together a coherent picture of cosmic history concerns studies of the
era corresponding to a redshift interval from 25 down to about 6; this corresponds to the period 200 million to 1 billion
years after the Big Bang. During this time the Universe apparently underwent two vitally important changes. Firstly, 
the earliest stellar systems began to shine, bathing the Universe in ultraviolet radiation from their hot, metal-free stars. 
Although isolated massive stars may have collapsed  and briefly shone earlier, the term {\it cosmic dawn} usually 
refers to the later arrival of dark matter halos capable of hosting star clusters or low mass galaxies. Secondly, the 
intergalactic medium transitioned from a neutral and molecular gas into one that is now fully ionized - a process 
termed {\it cosmic reionization}.

It is tempting to connect these two changes via a cause and effect as illustrated in Figure 1. Young stellar 
systems forming at a redshift of 25, corresponding to 200 Myr after the Big Bang, emit copious amounts of 
ultraviolet radiation capable of ionizing their surroundings. These ionized spherical bubbles expand with time
and, as more stellar systems develop, they overlap and the transition to a fully ionized intergalactic
medium is completed. 

\begin{figure*}
\centerline{\psfig{file=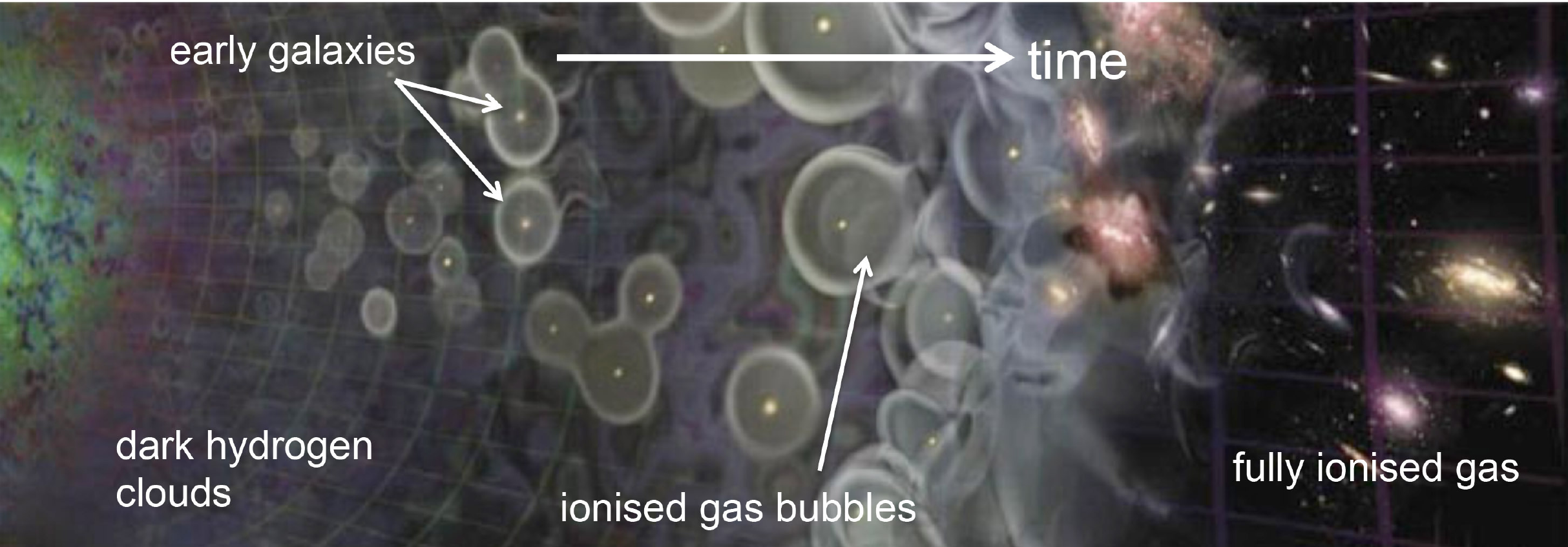,width=4.8in}}
\caption{An illustration of how early populations of star-forming galaxies reionized the Universe. Baryonic gas
is attracted into assembling dark matter halos and it cools and collapses to form the first stellar systems. Ultraviolet
radiation from their hot young stars photoionizes the surrounding neutral hydrogen creating ionized bubbles. As
more systems collapse and the ionized bubbles increase in size, their volumes overlap and cosmic reionization
is complete\cite{Loe06} .}
\label{rse:fig1}
\end{figure*}

In addition to determining when this transition occurred and whether this simple picture is correct,
studies of galaxies and the nature of the intergalactic medium during this period are valuable in
further ways. The relevant physical processes governing star formation at this time determine which 
primitive systems survive and which form the basic material for the subsequent evolution of galaxies. Indeed, relics of
this period may be present in local low mass dwarf galaxies devoid of star formation.
The abundance of the earliest low mass systems depends sensitively on the assembly history of
the dark matter halos which, in turn, depends on its streaming velocity. Although the cold
dark matter picture is favored by large scale structure observations, early galaxy formation would be delayed
if the dark matter was somewhat warmer and so direct observations of very early galaxies could verify or 
otherwise the standard picture\cite{Pac13}\ .

Ambitious facilities are now under construction, motivated in part by studies of the reionization era.
These include the James Webb Space Telescope (JWST) which has the unique capability to undertake
spectroscopy longward of 2$\mu$m, thereby accessing familiar rest-frame optical nebular
lines as measures of the ionizing radiation field and the evolution of the gas phase metallicity. Three
next-generation 25-40 meter aperture ground-based telescopes (the European Extremely Large Telescope, 
the Thirty Meter Telescope and the Giant Magellan Telescope) are also under development 
which will improve the spectroscopic capabilities. High order adaptive optics will
give these facilities impressive imaging capabilities, a highly relevant advantage as the faintest sources at
early epochs are otherwise unresolved. Deep near-infrared imaging over large areas of sky by survey facilities such as 
the European Space Agency's {\it Euclid} and NASA's WFIRST-AFTA missions will significantly improve information on the
demographics of early galaxies which is currently limited by cosmic variance uncertainties
associated with the small fields of view of the Hubble and Spitzer Space Telescopes.

These impressive upcoming facilities will be complemented by independent probes of the distribution
of cold and ionized gas charted tomograpically using the redshifted 21cm line.
Initial {\it pathfinder} projects such as the Low Frequency Radio Array (LOFAR) will address the statistical distribution
over a limited redshift range, whereas the Square Kilometer Array (SKA) will have the power to
directly image the evolving distribution of neutral gas. The combination of clustering statistics for the
early galaxy distribution and equivalent data for the neutral gas will delineate the evolution
of ionized regions in the context of the radiation from observed sources. This will revolutionize our
understanding of the reionization era.

In this brief review I take stock of what we currently know about the two principal questions that address
the picture illustrated above: when did reionization occur and were galaxies the primary reionizing agents? Although
we can address these questions using a variety of approaches, I will focus primarily on
what we are learning from studies of early star-forming galaxies. This naturally leads to a discussion
of the prospects for the next few years, including those possible with the future facilities listed above. Finally, I
list some of the fundamental challenges faced in interpreting the growing amount of data on early
galaxies.  My review is to be read in conjunction with a complementary discussion presented
by Steve Furlanetto in this volume which focuses more on
the theoretical aspects of reionization and the future prospects with 21cm tomography.

\section{When Did Reionization Occur?}

The earliest constraints on the reionization history arose from the Gunn-Peterson test\cite{Gun65}
applied to the absorption line spectra of $z>5.5$ QSOs (see [\refcite{Fan06}]).
The decreasing transmission due to thickening of the Lyman alpha forest was initially used to
argue that the reionization process ended at a redshift close to 6. However, only a very
small change in the volume-averaged fraction of neutral hydrogen, $x(HI)\simeq10^{-3}$,
is required to completely suppress the spectroscopic signal shortward of Lyman alpha
in the spectrum of a QSO, above which saturation rapidly occurs. Accordingly, this method
is only useful for detecting a subtle change at the end of the reionization process. Since
the bulk of the high redshift QSOs were analyzed some 8-10 years ago\cite{Fan02,Fan06},
progress in locating higher redshift QSOs has been slow. Fortunately, some additional constraints have
been provided through equivalent spectroscopy of a handful of  $z\gtrsim$6 long duration gamma ray burst 
(GRB) afterglows\cite{Cho14}\ . Unfortunately, none of the more distant GRBs discovered beyond $z\simeq$7 
was followed up in detail. Indeed, only one source above a redshift of 7 - a QSO - has a relevant absorption 
line spectrum above a redshift of 7 [\refcite{Mor11}]\ . The initial analysis of this spectrum suggested that the IGM
may indeed be significantly neutral ($x(HI)\simeq10^{-1}$) at this redshift ([\refcite{Mor11,Bol11}] but
see Boseman \& Becker, in prep.), although confirmation from additional lines of sight is clearly desired.

A second constraint on the reionization history arises from the optical depth $\tau$ to
electron scattering to cosmic microwave background (CMB) photons and the cross-correlation 
of the polarization signal induced by these electrons and the temperature fluctuations. 
$\tau$ therefore acts as a integral constraint on the line of sight distribution of ionized gas.
The angular correlation can be interpreted in structure formation theory as providing an approximate redshift
of the reionization era. Usually the quoted result corresponds to that assuming an (unrealistic) an
instantaneous reionization. Over the past few years WMAP has provided a series of improved constraints\cite{Hin13}
corresponding to instantaneous reionization at $z\simeq10.6 \pm 1.1$.
No polarization results are yet available from Planck mission but early constraints
based on temperature fluctuations alone\cite{Pla13} are consistent. It will be
very important to secure independent confirmation of $\tau$ from the Planck mission.
The prospects of using higher order CMB data to improve our understanding of reionization 
in the future is discussed by Calabrese et al\cite{Cal14}\ .

The most recent development in tracing reionization history follows studies of the rate of occurrence
of Lyman alpha (Ly$\alpha$) emission in star-forming galaxies. Miralda-Escud\'e\cite{Mir98}
and Santos\cite{San04} discussed the prospect of using Ly$\alpha$ as a resonant transition,
one which is readily absorbed if a line emitting galaxy lies in a neutral IGM.
Early results based on the luminosity functions of narrow-band selected Ly$\alpha$ emitting
galaxies over the redshift range $5.7<z<6.5$ supported the notion of a rapidly-changing IGM
via a marked decline in the abundance of emitters over a short period
of cosmic history (corresponding to an interval of less than 200 Myr)\cite{Kas06,Ouc10}\ . However, although a striking
result, it is hard to separate the effect of an increasingly neutral IGM at high redshift from the 
declining abundance of star-forming galaxies deduced from the overall population observed
beyond $z\simeq$4 [\refcite{Bou10}]\ .

An improved test that removes this ambiguity involves measuring
the {\it fraction} of line emission in well-controlled, color-selected
Lyman break galaxies. First introduced as a practical proposition by 
Stark\cite{Sta10}\, this method has been variously applied in the last 
3 years\cite{Pen11,Sch12,Tre13}\, and most recently, by 
Schenker et al\cite{Sch14}\ . The availability 
of large numbers of $z>7$ candidates from deep HST imaging and new multi-object
near-infrared spectrographs has enabled considerable progress of late.
These observations confirm a marked decline in the visibility
of Ly$\alpha$ beyond a redshift $z\simeq$6.5, consistent with the
Gunn-Peterson constraints discussed above (Figure 2). Although Schenker et al
report spectroscopic data for 102 $z>6.5$ Lyman break galaxies,
only a handful beyond $z\simeq$7 show Ly$\alpha$ emission, the current 
record-holder being at $z=7.62$. 

The challenge lies in interpreting the fairly robust decline in the visibility
of Ly$\alpha$ emission in the context of an increasing neutral fraction 
$x(HI)$ at earlier times. Radiative transfer calculations have
suggested the fast decline in Figure 2 could imply a 50\% neutral
fraction by volume as late as $z\simeq$7.5\cite{Mes14,Smi14}\ .
The uncertainties in this interpretation include (i) cosmic
variance given the limited volumes so far probed with ground-based
spectrographs\cite{Tay14}\ , (ii) the assumed velocity offset of 
Ly$\alpha$ with respect to the systemic velocity of the galaxy which
is critical in understanding whether the line resonates with any neutral
gas\cite{Sch13a,Sta14} and (iii) the possible presence
of optically-thick absorbing clouds within the ionized regions\cite{Bol11}\ . 
A final variable is the escape fraction of ionizing photons from
the galaxy, $f_{esc}$. If this were much higher at earlier times
as a result of less neutral gas in the galaxies, the production of
Ly$\alpha$ in the intrinsic spectrum would be reduced\cite{Dij14}\ .

\begin{figure*}
\hbox{
\psfig{file=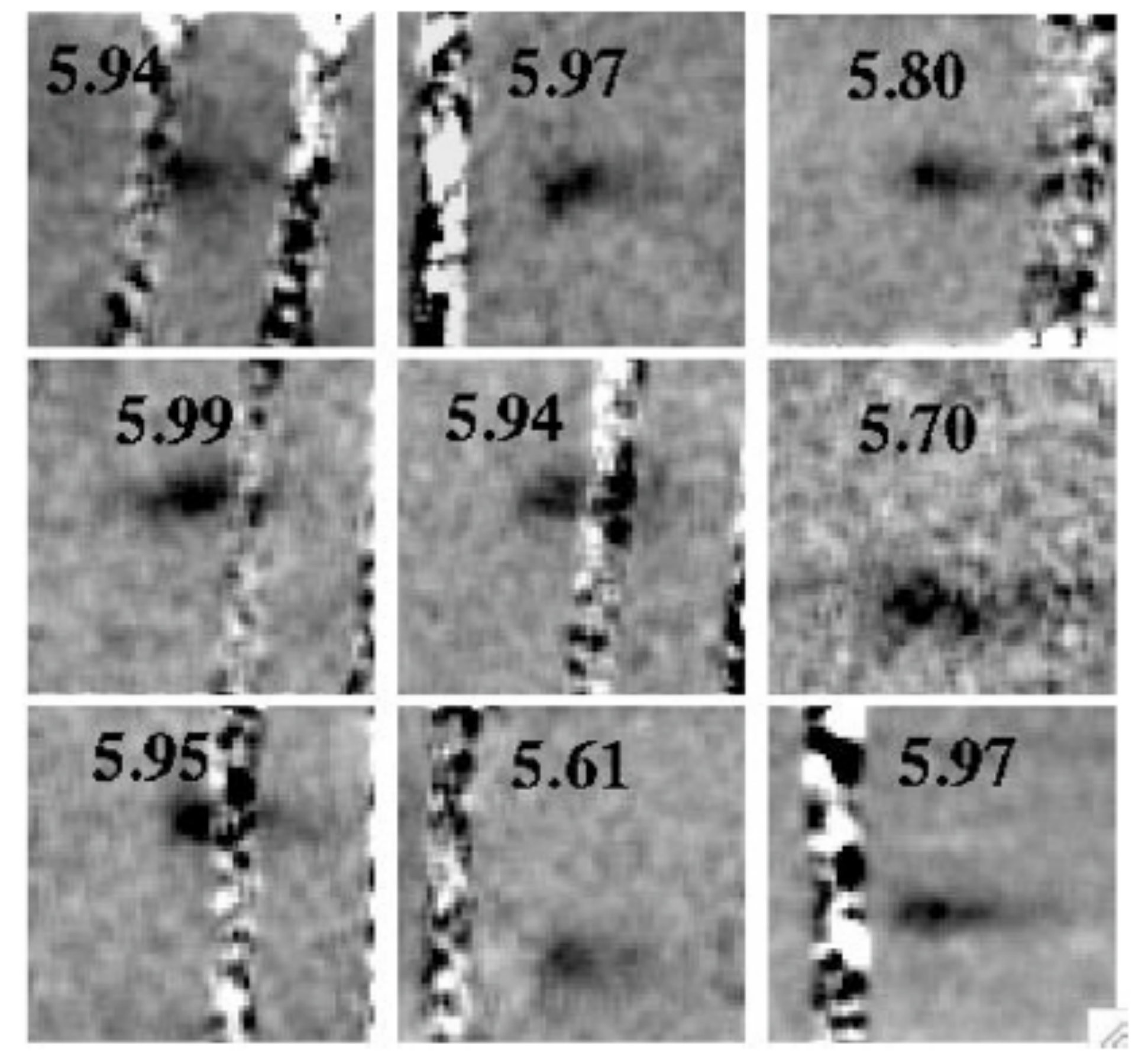,width=2.0in}
\psfig{file=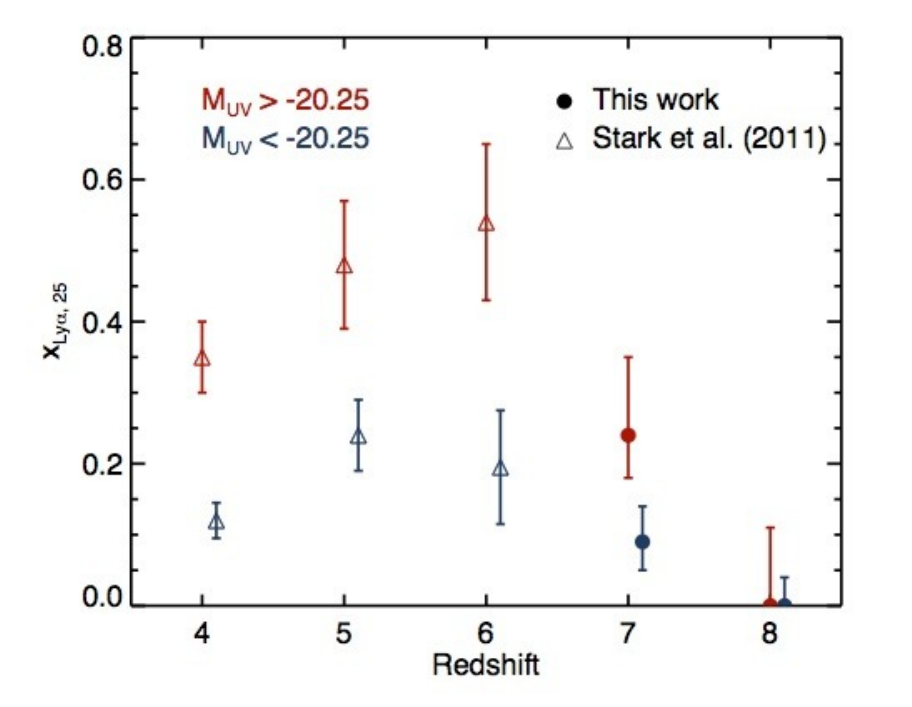,width=3.0in}
}
\caption{{\bf Left:} Keck spectra for $z\simeq$6 galaxies showing the high rate
of occurrence of Ly$\alpha$ emission. Each sub-panel represents a 2-D spectrum
of a Lyman break galaxy and the black regions represent line emission
at the redshift marked\cite{Sta10}\ .  {\bf Right:} The evolving fraction of Lyman break galaxies 
in various luminosity bins that present a detectable Ly$\alpha$ emission line from the recent survey of
Schenker et al\cite{Sch14}\ .  The rising fraction over $4<z<6$ is interpreted
via a reduced dust extinction at early times, whereas the sudden reversal
beyond $z\simeq$6 is attributed to an increasingly neutral intergalactic medium. }
\label{rse:fig2}
\end{figure*}

A complementary and promising method for tracing reionization is
to statistically chart the evolving distribution of neutral gas directly via
redshifted 21cm emission using radio interferometers such as
LOFAR\cite{Mel13} and the Murchison Wide Field
Array\cite{Bow13}\ . No direct detections are yet available but the
prospects are discussed by Steve Furlanetto elsewhere in this volume.

Figure 3 represents a recent summary of the various constraints
on reionization and includes several methods not described
in this brief review\cite{Rob13} . As can be seen, the
redshift range 6 to 20, corresponding to a period of 800 Myr is 
considered to be the window of interest. 

\begin{figure*}
\centerline{\psfig{file=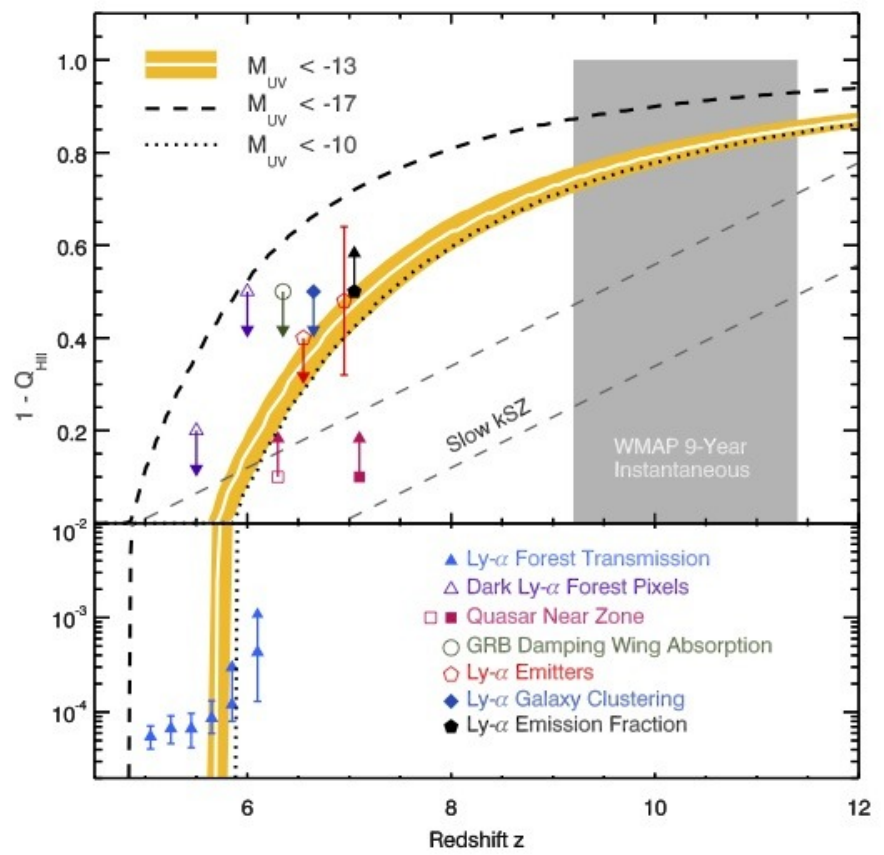,width=3.8in}}
\caption{Reionization histories for models that include galaxies to various
luminosity limits from the UDF12 survey ($M_{UV} < 13$ white line; 68\% 
credibility region: orange area; $< 17$, dashed line; $< 10$, dotted line) plus 
other claimed constraints on the neutral fraction  $1 - Q_{HII}$ (see lower panel
for legend). Methods not discussed in the text include the fraction of dark pixels in the 
Ly$\alpha$ forest (purple open triangles), QSO near-zone measurements (open and solid 
magenta squares), damping wing absorption in a GRB (open green circle),  
the clustering of Ly$\alpha$ emitters (filled dark blue diamond). The gray dashed 
lines labeled ÒSlow kSZÓ illustrate the slowest evolution permitted by small-scale CMB 
temperature data and the shaded gray region shows the redshift of instantaneous 
reionization according to WMAP\cite{Rob13} . }
\label{rse:fig3}
\end{figure*}

\section{Were Galaxies Responsible for Cosmic Reionization?}

Potential contributors to the reionizing photons include star-forming galaxies, 
non-thermal sources such as quasars and low luminosity active galactic
nuclei, primordial black holes and decaying particles.
Luminous QSOs decline rapidly in their abundance beyond $z\simeq$6 
so the only prospect for non-thermal sources contributing significantly
to reionization might be if the faint end of their luminosity function is unusually steep.
Current estimates of the high redshift AGN luminosity function suggest
this is not the case although the observational uncertainties are 
still large\cite{Gli11,McG13}\ .

Star-forming galaxies represent the most promising reionizing source 
given they are now observed in abundance in the relevant redshift
range from deep surveys such as the Hubble Ultra Deep Field (UDF)\cite{Koe13}\ .  
These and other data reveal a steep luminosity function at the faint end\cite{Sch13b,McL13,Fin14}\  ,
such that it is reasonable to assume we are only observing the luminous fraction
of a much larger population. However, a quantitative calculation of the photon budget 
requirements for maintaining reionization involves additional parameters, some 
of which are largely unconstrained (see recent review by [\refcite{Rob13}]).

In this case, the reionization process is a balance between the recombination
of free electrons with protons to form neutral hydrogen and
the ionization of hydrogen by Lyman continuum photons. The
dimensionless volume filling factor of ionized hydrogen $Q_{HII}$
can be expressed as a time-dependent differential equation:

$$\dot{Q}_{HII} = \frac{\dot{n}_{ion}}{<n_H>} - \frac{Q_{HII}}{t_{rec}}$$

The recombination time $t_{rec}$ depends on the baryon density, the
primordial mass fraction of hydrogen, the case B recombination coefficient
and the clumping factor $C_{HII} \equiv <n_H^2>/<n_H>^2$ which takes
into account the effects of IGM inhomogeneity through the quadratic
dependence of recombination on density.  Simulations suggest
$C_{HII}\simeq$1-6 at the relevant redshifts\cite{Fin12}\ ,
although there has been much discussion of its redshift dependence
depending on the epoch when the ultraviolet (UV) background becomes uniform.
If the clumping factor $C_{HII}$ is time invariant, $t_{rec}$ declines with increasing
redshift. For the expected values above, at redshifts $z<10$, $t_{rec}$ 
exceeds 100-200 Myr\cite{Paw09,Rob10} ensuring
recombination is unlikely. However, if the source of ionizing photons
is not steady in the redshift range $10<z<25$, there remains the possibility
of an intermediate recombination era, perhaps inbetween reionization 
from the first isolated massive stars and that subsequently from early galaxies.

The main uncertainty in understanding the contribution of galaxies
can be understood via the relative contributions to the ionizing
photon rate $\dot{n_{ion}}$:

$${\dot{n}_{ion}} = f_{esc} \xi_{ion} \rho_{SFR}$$

where $\rho_{SFR}$ represents the most direct observable, the
integrated volume density of star-forming galaxies. This involves
measuring the redshift-dependent luminosity function, typically
in the rest-frame UV continuum ($\simeq$ 1500 \AA\ ) which 
is accessible at $z\simeq7-10$ with HST's near-infrared camera WFC3/IR,
and above $z\simeq$10 with NIRCam on JWST.
The faint end slope of the luminosity function is a critical factor given it contributes
the major portion of the integrated luminosity density\cite{Sch13b,McL13,Fin14} . 
$\xi_{ion}$ is the ionizing photon production rate which encodes
the number of photons more energetic than 13.6 eV that are produced
per unit star formation rate.  This requires knowledge of the stellar
population which can currently only be estimated by modeling
the average galaxy color. Finally, $f_{esc}$ represents the fraction
of ionizing photons below the Lyman limit which escape to the IGM.
This is the least well-understood parameter. It can only be directly
evaluated through rest-frame UV imaging or spectroscopy
at $z\simeq2-3 ([$\refcite{Sia10,Nes11}]) where values
as low as 5\% are typical. At higher redshift, any photons
below the Lyman limit are obscured along the line of sight
by the lower redshift Lyman alpha forest.

There are several ways to address the question of whether
galaxies can meet the ionization budget and these depend
critically on the assumed value of the currently unobserved quantities, e.g. $f_{esc}$. 
A fundamental requirement is that the integrated electron
path length to the start of reionization should match the
optical depth of Thomson scattering, $\tau$, in the CMB. When this
requirement is imposed, in the context of the results from the Hubble
UDF, three conditions are necessary for galaxies to be the main reionization
agents\cite{Rob13}\ . Firstly, the escape fraction
$f_{esc}$ has to rise with redshift or be sufficiently luminosity-dependent 
so that at least 20\% on average of the photons escape a 
typical low luminosity $z\simeq7-10$ galaxy. Secondly, galaxies must populate the luminosity function
to absolute magnitudes below the limits of the deepest current HST images at 
$z\simeq7-8$ ($M_{UV} = -17$). Finally, the galaxy population must extend 
beyond a redshift $z\simeq10$ to provide a sustained source of ionizing radiation.
Various combinations of these three requirements have been
discussed in the literature and presented alternatively as reasonable
assumptions or as critical shortfalls in the ionizing budget!

\begin{figure*}
\hbox{
\psfig{file=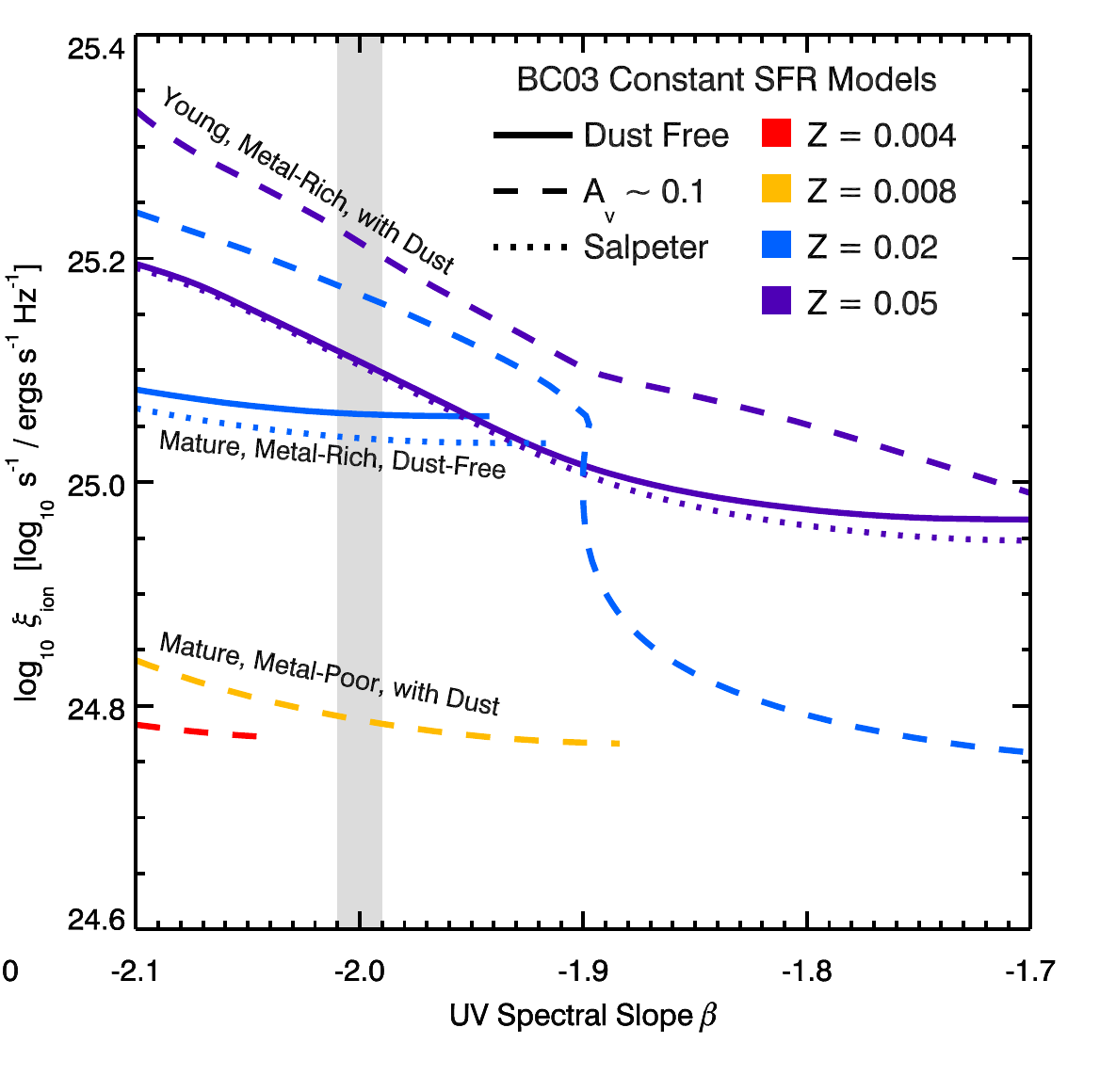,width=2.5in}
\psfig{file=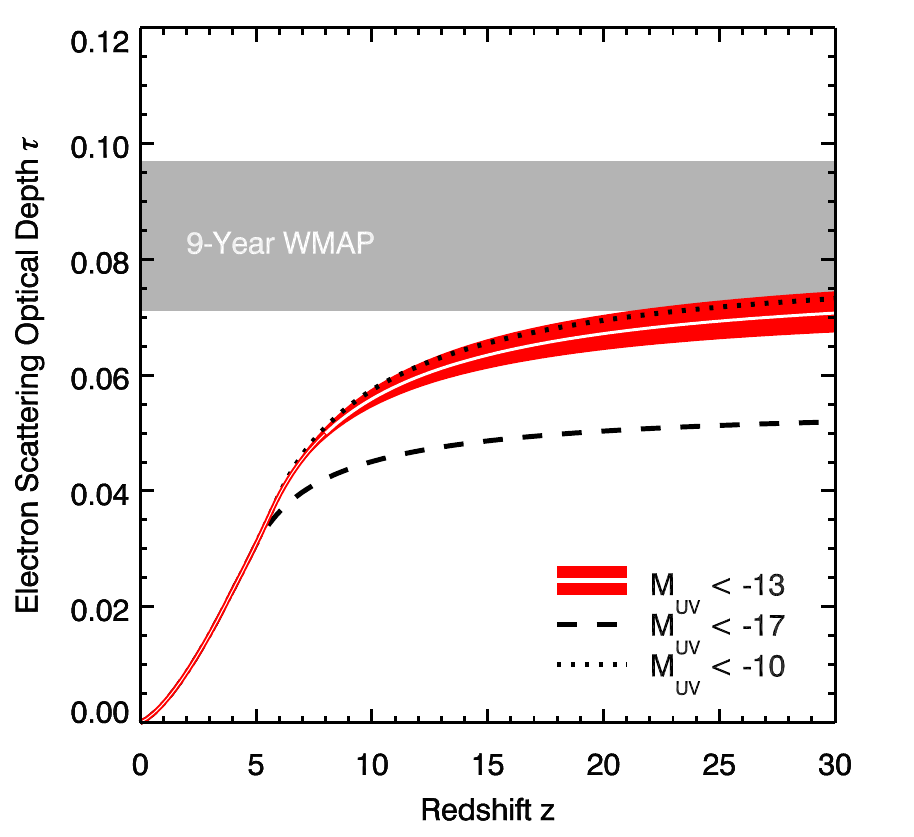,width=2.5in}}
\caption{{\bf Left:} Degeneracies in inferring the ionizing photon 
production factor $\xi_{ion}$ in terms of the observed slope $\beta$
of the ultraviolet continuum, the gray shaded area being that
observed for $z\simeq7-8$ galaxies\cite{Dun13} . Time
tracks are shown for stellar population synthesis models of varying
dust content, metallicity and the initial mass function\cite{Rob13} . 
{\bf Right:} One aspect of the UV `photon shortfall' for galaxies as agents
of reionization given the abundance of galaxies in the UDF.
Assuming a 20\% escape fraction and continuity in the declining
star formation rate density beyond $z\simeq10$, the figure shows
the need to extend the UV luminosity function lower than
the current $M_{UV}=-17$ detection limit to reproduce the optical
depth of electron scattering in the WMAP data\cite{Rob13} .}
\label{rse:fig4}
\end{figure*}

A further constraint on the above is the requirement that
the sum of the star formation during the reionization era
cannot exceed the stellar mass density observed using
the Spitzer satellite at the end of reionization, say $z\simeq5-6$ ([\refcite{Sta07}]).
This mid-infrared satellite is uniquely effective in this regard given its
infrared camera, IRAC, surveys high redshift galaxies
at rest-frame optical wavelengths where longer-lived
stars can be accounted for. Formally, this can be expressed:

$$\rho_{\ast} (z=6)  = C \, \int_{z=6}^{\infty} \int \Phi (L_{UV}, z) L_{UV} dL dz$$

where $\rho_{\ast}$ is the required stellar mass density per comoving
volume at the end of reionization, and $C$ represents the necessary 
factor to convert the observed redshift-dependent UV luminosity 
function $\Phi (L_{UV})$ and its associated luminosity density, into
a star formation rate density. Stellar masses for individual galaxies
are usually determined by deriving a mass/light ratio from fitting
the spectral energy distribution and multiplying by the luminosity.
To secure the integrated mass density is challenging given only
a more massive subset of the $z\simeq$6 population is currently
detectable with Spitzer. Additionally the Spitzer photometric
bands are likely contaminated by nebular line emission at
$z\simeq$6 and significant, but uncertain, downward corrections are
required to estimate the true mass density\cite{Sta13,Sch13a}. When
reasonable estimates are made of the unseen stellar mass
at $z\simeq$5-6 and corrections applied for nebular emission
based on spectroscopic evidence at lower redshift, the stellar
mass densities $\rho_{\ast}$ can be reconciled with the earlier
star formation history\cite{Rob13}\ .

\section{The Near Future}

Fortunately we observers have not yet reached a threshold
in exploring the early galaxy population pending the arrival of
new facilities such as JWST and the next generation of large
ground-based telescopes. There are several interesting and immediate
initiatives available for making further progress.

In addition to probing the reionization history with the fractional rate
of occurrence of Ly$\alpha$ emission, the spatial distribution of
line emitters in principle contains data on the topology of ionized
regions where emission can be transmitted. Narrow-band filters
are being used with panoramic cameras to locate Ly$\alpha$ emitters
at discrete redshifts where the line is favorably placed with respect
to the night sky emission, for example at redshifts $z$=5.7, 6.6 and 7.1
with the HyperSuprime-Cam 1.5 degree field imager on the Subaru 
8.2m telescope (see an example of earlier work of this nature in Figure 5). 
The correlation of such line emission with redshifted 21cm emission would
be a particularly fruitful program.

\begin{figure*}
\centerline{\psfig{file=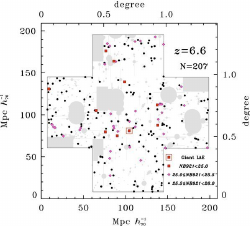,width=3.5in}}
\caption{Angular distribution of 207 Ly$\alpha$ emitters at a redshift of 
$z$=6.565 $\pm$ 0.054 selected from a mosaic of narrow band images taken with
the Suprime Camera on the 8.2m Subaru telescope, color coded according
to their luminosity (decreasing from red squares, through magenta diamonds
to black circles\cite{Ouc10}\ . The open red square denotes the extended and luminous
emitter `Himiko' (see Section 5). }
\label{rse:fig5}
\end{figure*}

Strong gravitational lensing by foreground clusters offers a
valuable tool for exploring the redshift range $7<z<10$ population.
HST and Spitzer are investing significant resources in deep imaging
of selected clusters via the CLASH \cite{Pos12} and Frontier
Fields\footnote{\rm http://frontierfields.org} programs. 
Lensing facilitates two broad applications depending on the
source magnification involved. Bradley et al\cite{Bra14} discuss the 
magnification distribution for the CLASH survey and Richard et al\cite{Ric14} 
for the upcoming Frontier Field clusters. Most of the lensed sources 
have magnifications of $\times$1.5-3 with less than 5\% greater than $\times$10
(Figure 6a).

The first regime involves very highly-magnified and usually multiply-imaged sources 
observed close to the critical line of the cluster. With magnifications of 
$\times10-30$\cite{Ell01,Kne04,Coe13,Zit14} such systems offer the prospect of 
valuable detailed studies. A good example is the $z\simeq$6.02 galaxy in the rich cluster 
Abell 383 which has a magnification of $\times11.4\pm1.6$
corresponding to a 0.4\,$L^{\ast}$ galaxy\cite{Ric11}\ .
The significant boost in brightness enables a much more precise
spectral energy distribution for a representative sub-luminous
system than would otherwise be the case providing a fairly robust
stellar age of 640- 940 Myr, corresponding to a formation redshift of 
$z\gtrsim15$. However, such configurations are rare and do not
represent a straightforward route to large samples.

\begin{figure*}
\hbox{
\psfig{file=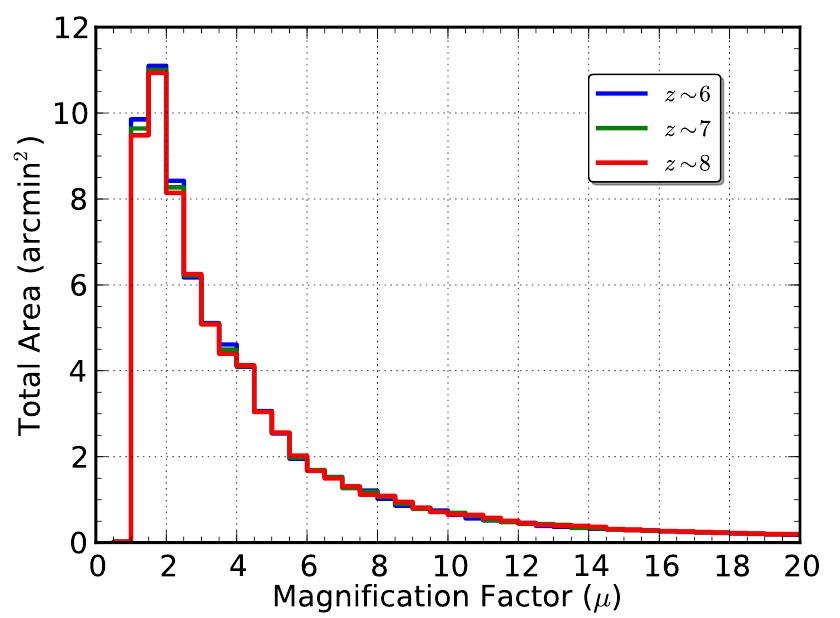,width=2.6in}
\psfig{file=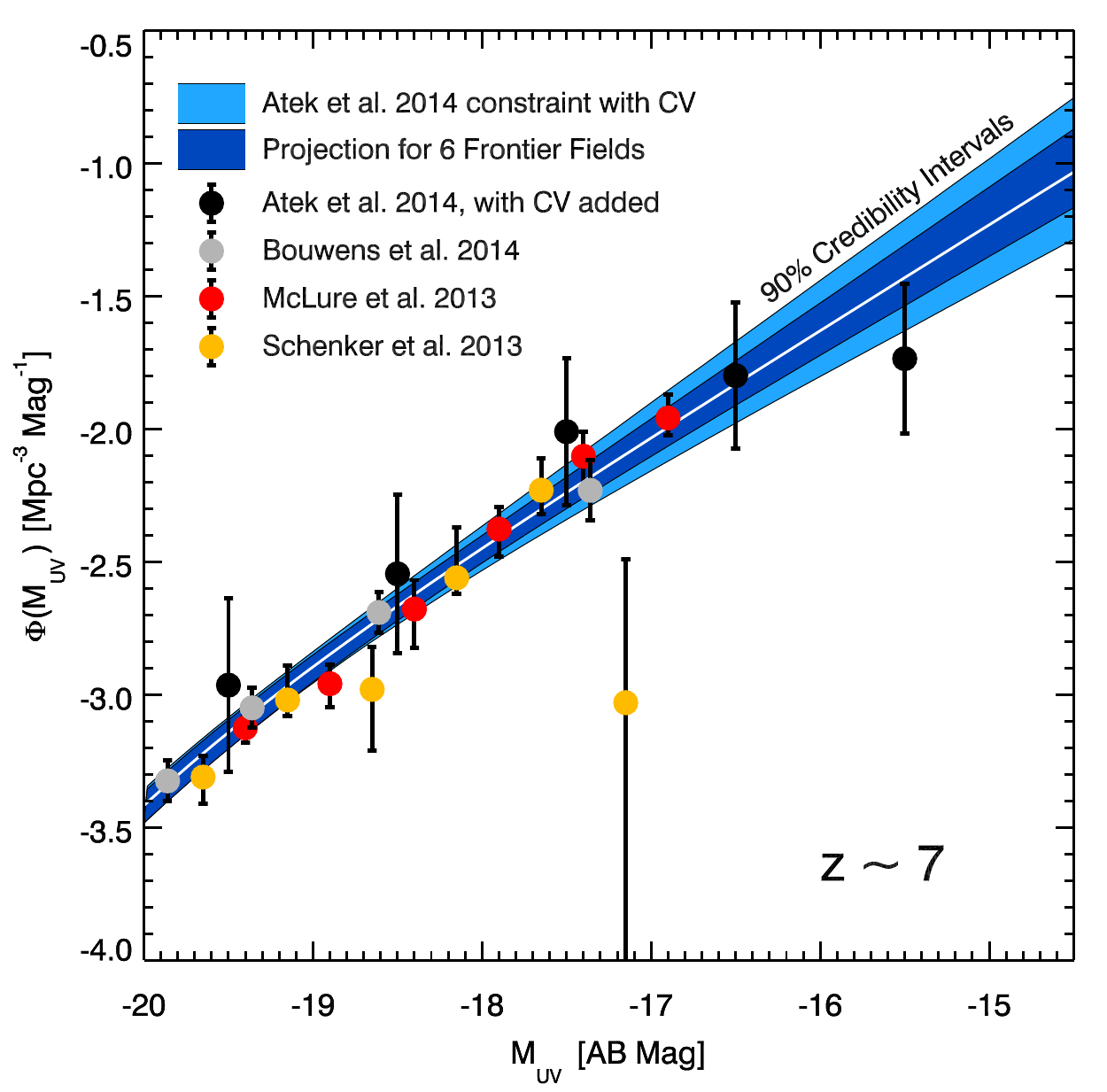,width=2.2in}
}
\caption{{\bf Left:} The distribution of lensing magnifications deduced
for high redshift galaxies in the HST CLASH survey of foreground clusters\cite{Bra14}.
Most sources are only modestly magnified but there are rare sources
magnified by factor as large as $\times$10-20.
{\bf Right:} Using gravitational lensing in the HST Frontier Field 
program to extend the $z\simeq$7 luminosity function fainter
than was possible in the UDF. Using the available data for one
cluster\cite{Ate14}, a projection is made for all 6 clusters\cite{Rob14}. }
\label{rse:fig6}
\end{figure*}

The second regime involves more modest magnifications of larger numbers
of background sources. The benefits here are not in detailed studies of
individual sources but rather for statistical purposes, e.g. in extending the 
$z\simeq$7-8 luminosity function fainter than was possible in the deepest 
blank field studies\cite{Ate14} (Figure 6b). Robertson et al\cite{Rob14} recently 
projected the likely gain in depth over all 6 Frontier Field clusters incorporating
the increased cosmic variance in lensed surveys. They claim the uncertainty in the 
faint end slope $\alpha$ of the luminosity function would be significantly
reduced compared to the value in the UDF ($\Delta\alpha=\pm 0.05 $ c.f. $\pm$ 0.18).

Detailed spectroscopy of $z\simeq7-8$ galaxies can also provide 
further information on the ionization state and metallicity of the gas.
Stark et al\cite{Sta14} illustrate how, even when Ly$\alpha$ is suppressed
by neutral gas, other nebular lines such as CIII] 1909 and CIV 1550 \AA\
are within reach of current near-infrared spectrographs, although this
is highly challenging work even for lensed sources.

This leads naturally to the longer term goal of gathering {\it gas-phase metallicities}
for early galaxies thereby adding {\it chemical enrichment} as the next logical
tracer of earlier activity. Metallicity measurements will very much be the province of JWST given
all the familiar rest-frame optical lines ([O II], [O III], H$\alpha$, [N II])
used locally and at intermediate redshifts as well-calibrated metallicity
indicators, are shifted beyond 2$\mu$m where ground-based spectroscopy 
of faint objects is impractical. However, there are valuable sub-mm lines 
accessible with ALMA at high redshift which may give information on
both the metallicity and dust content of early galaxies. Although the 
currently-held view is that the blue UV colors of most of the $z>7$ galaxies implies
little or no dust, strong ALMA upper limits on far-infrared continua would provide
a more convincing argument.

The [CII] 158 $\mu$m line has traditionally been one of the most
valuable tracers of star formation in energetic sources and a correlation 
is often claimed between the [C II] luminosity and the star formation
rate estimated from the far infrared flux although its interpretation 
remains unclear\cite{deL11}\ .  Early ALMA studies of 
luminous $z\simeq5-7$ dusty starbursts recovered prominent
[CII] emission\cite{Rie13,Rie14} consistent with
this correlation. However, an intense Ly$\alpha$ 
emitter, dubbed `Himiko' at $z$=6.595 (see Figure 5) with a high star formation 
rate ($\simeq100 M_{\odot}$ yr$^{-1}$) reveals no far infrared
or [CII] emission\cite{Ouc13}\ , and thus deviates significantly
from the normal relation. As the Ly$\alpha$ emission is
particularly extended and the source is unusually luminous
compared to its cohorts, conceivably it is being observed
during a special moment in its history e,g. an energetic burst
of early activity in a very low metallicity system. Such studies
with ALMA may shed light on metal formation in the most luminous 
early systems ahead of the launch of JWST.

Ultimately one might hope to identify systems
with minimal pollution from metals. Such `Population III' sources
initially represented something of a `Holy Grail' for the next generation
facilities - specifically, the charge to find a star-forming galaxy
or stellar system devoid of metals. More recent numerical
simulations\cite{Wis12} indicate the self-enrichment
of halos from early supernovae is surprisingly rapid ($<$100 Myr)
and so such primordial `first generation' stellar systems may be
very rare.

\section{Outstanding Issues}

Although there a gaps in our quantitative knowledge of
the reionization history and the role of galaxies, it has perhaps
become commonplace to regard sketched histories such
as Figures 3 and 4b as the correct framework within which
future facilities can fill in the details. In this concluding
section I want to highlight some outstanding issues and
puzzles that will serve to focus our collective research
in the near future.

\noindent{\bf The extent of star formation beyond $z\simeq$10:}
The Ultra Deep Field 2012 campaign argued for a 
near-continuous decline in the cosmic star formation rate 
density over $4<z<10$ (Ref [\refcite{Ell13}]) and Robertson
et al\cite{Rob13} used this continuity plus the mature
ages of the $z\simeq$7-8 galaxies\cite{Dun13}\ ,
as indirect evidence that the star formation history
beyond $z\simeq$10. However, recent work exploiting
the wider, but shallower CANDELS data\cite{Oes14}
together with several analyses exploiting early Frontier
Field lens data\cite{Ish14} point to a discontinuity in this
decline at $z\simeq$8. Such a downturn would be hard to
reconcile with the stellar mass density evolution\cite{Ric11,Lab13} 
and, if correct, would seriously increase the UV photon budget shortfall.
A key issue here, given the paucity of data beyond $z\simeq$8, is 
uncertainties arising from cosmic variance\cite{Rob14}\ . 
Hopefully with further data from the Frontier Fields and more
Spitzer age measures of individual galaxies at $z\simeq$7-8, the 
situation will be clarified ahead of the launch of JWST.

\noindent{\bf Missing star-forming galaxies:} The high redshift
galaxies discussed in this review have almost
exclusively been located by their ultraviolet emission, either
via continuum colors or through Ly$\alpha$ emission.
In addition to assuming there are yet fainter galaxies
further down the luminosity function beyond HST's
limits, is it conceivable there are additional sources
perhaps dusty or those not selected via the current
methods? An unresolved puzzle is the anomalously
high rate of long duration gamma ray bursts seen
beyond $z\simeq$5 compared to that expected
using a GRB rate normalized to the star formation rate
observed at lower redshift\cite{Rob12} (Figure 7a).
This discrepancy may be telling us more about
the evolving production rate of GRBs in low
metallicity environment rather than something
fundamental about the cosmic star formation history.
Nonetheless, it acts as a warning that some aspects
of early massive star formation may not be understood.

\begin{figure*}
\hbox{
\psfig{file=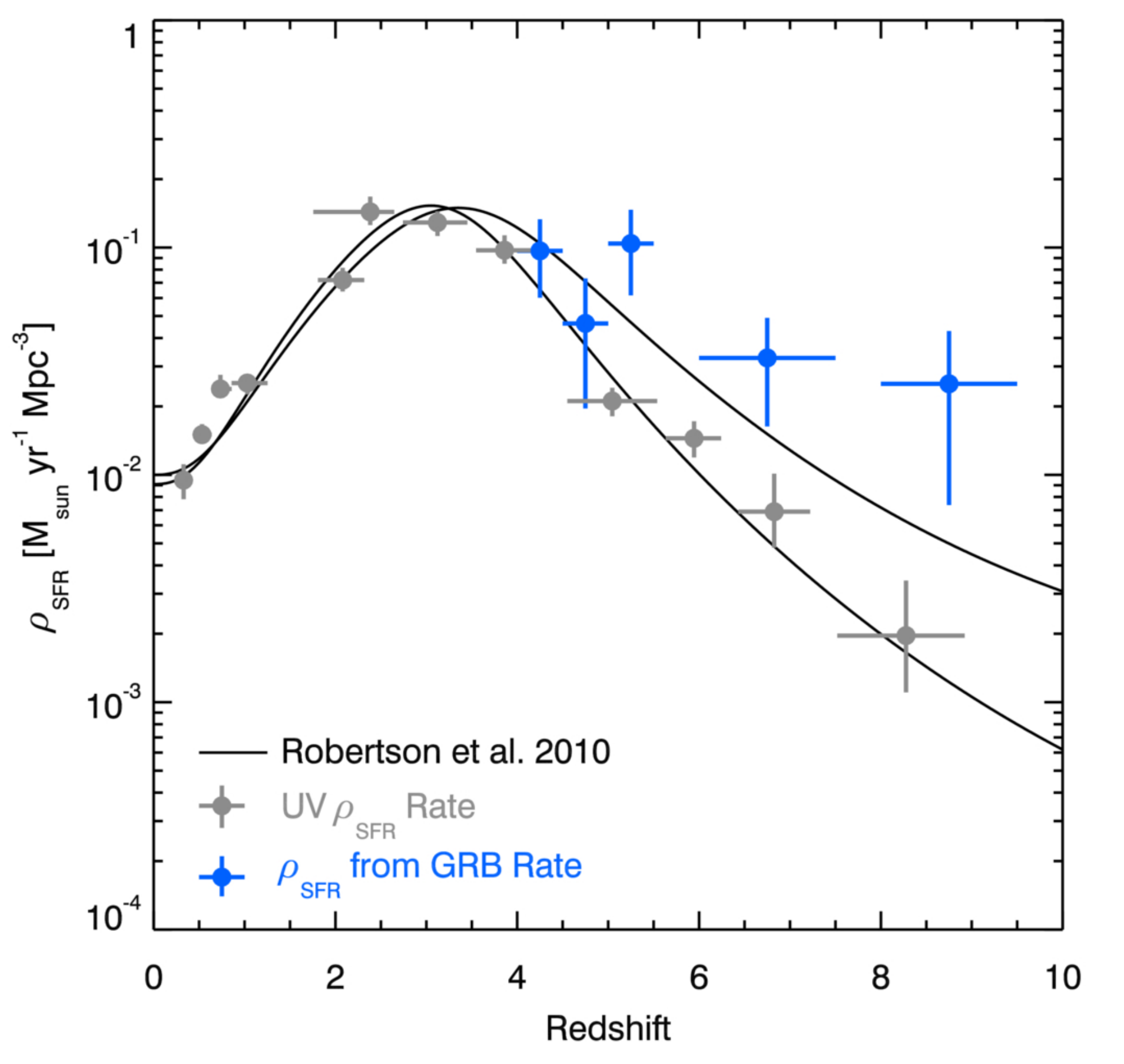,width=2.2in}
\psfig{file=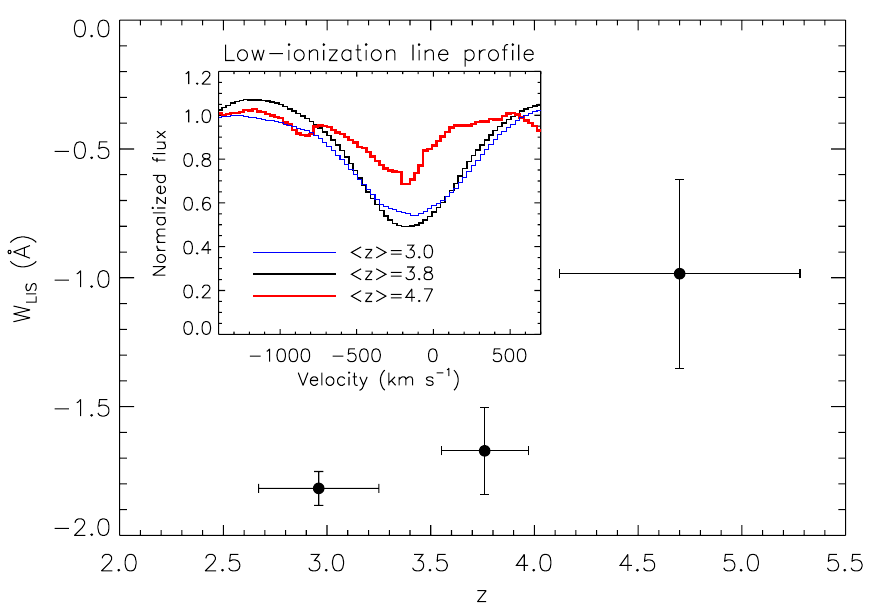,width=2.8in}
}
\caption{{\bf Left:} The star formation history beyond
a redshift 4 as inferred from the rate of gamma ray
bursts (GRBs)\cite{Rob12} . The number of GRBs is
converted into a volume-averaged star formation by
matching their cumulative redshift distribution over
$0<z<4$ with the cosmic star formation history. There
is a worrying excess in the number of high z GRBs compared
to expectations based on the rate at lower redshift.
{\bf Right:} The equivalent width of low ionization
gaseous absorption lines from Keck spectra of Lyman
break galaxies stacked at various redshifts. The inset 
shows the stacked absorption line profiles whose depth 
becomes shallower as the redshift increases. Such
data suggests that the covering fraction of neutral
gas is less at high redshift and hence the escape
fraction increases\cite{Jon12} . }
\label{rse:fig7}
\end{figure*}

\noindent{\bf The escape fraction of ionizing photons:} The
largest uncertainty in addressing the role of galaxies
in completing the reionization process is the
average fraction of ionizing photons that can escape
a typical low-luminosity galaxy. Even with a fraction $f_{esc}\simeq$,
20\% there is significant tension in the ionizing budget
and in reproducing the optical depth $\tau$ of electron
scattering by the CMB (Figure 4b). Most likely the escape fraction
varies significantly from galaxy to galaxy according to the
geometric viewing angle, kinematic state, star formation rate and 
physical size of each galaxy. Even at redshifts $z\simeq$2-3, determining
$f_{esc}$ has been a challenging endeavor although the 
consensus points to the range 0-5\%\cite{Sia10,Nes11}\ . 
At high redshift, the only practical route is to
examine the {\it covering fraction}, $f_{cov}$, of
neutral or low ionization gas on the assumption
that, typically, $f_{esc} = 1 - f_{cov}$. Even so, measuring $f_{cov}$
requires high signal/noise absorption line spectroscopy which is 
only practical for stacks of galaxies\cite{Jon12} or 
strongly-lensed examples\cite{Jon13}\ .
Such data to $z\simeq$4-5 shows some evidence for a rising escape
fraction with increased redshift (Figure 7b) but the method
needs to be extended to larger samples at yet
higher redshifts.

\noindent{\bf When did the Universe produce dust?}
To these more immediate issues of observational
interpretation should be added the question of
whether dust is present beyond $z\simeq$7. Its
presence would seriously confuse interpretations
of the UV colors (e.g. Figure 4a) as well as
raise the question of obscured star formation.
An example has recently been found of a
convincing ALMA continuum detection for a
star-forming galaxy at $z$=7.58 (Watson et al in prep)
which raises very interesting consequences.
This early result highlights the key role that
ALMA can play in complementing studies
of high redshift galaxies with HST and Spitzer.

\section*{Summary}

Although many puzzles remain as indicated above, the pace
of observational discovery is truly impressive and
will continue as we see the first convincing
results from 21cm interferometry in the next 1-2 years,
launch JWST in 2018 and commission the next
generations telescopes in the early 2020's.
The observational promise is evident and I
encourage our theoretical colleagues to get
ready for the next revolution in observational
data at the redshift frontier!

\section*{Acknowledgements}

I acknowledge valuable discussions with my co-Rapporteur,
Steve Furnaletto, a colleague on the Hubble
Ultra Deep Field (UDF) campaign. I likewise acknowledge
the support and scientific input from my other UDF colleagues,
Brant Robertson, Jim Dunlop, Ross McLure and Anton Koekemoer,
as well as my Keck spectroscopic co-workers Dan Stark,
Matt Schenker, Tucker Jones and Adi Zitrin. I thank the organizers of this
memorable meeting for their organizational efforts and
hospitality in Brussels.

\bibliographystyle{ws-procs975x65}
\bibliography{rse}

\end{document}